%% file: main.tex
\documentclass{article}
\usepackage{spconf,amsmath,graphicx}
\usepackage{float}
\usepackage{multirow}
\usepackage{booktabs}
\usepackage{hyperref}
\usepackage[dvipsnames]{xcolor}

\title{Cross-speaker encoding network for multi-talker speech recognition}
\name{Jiawen Kang, Lingwei Meng, Mingyu Cui, Haohan Guo, Xixin Wu, Xunying Liu, Helen Meng}
\address{The Chinese University of Hong Kong, Hong Kong SAR, China}
\begin{document}
\ninept
\maketitle
\begin{abstract}
\input{content/0-abstract}

\end{abstract}
%


\section{Introduction}
\label{sec:intro}
\input{content/1-intro}

\section{Cross-Speaker Encoding Network}
\label{sec:model}

\input{content/2-method}

\section{Experimental setup}
\label{sec:setting}
\input{content/3-settings}

\section{Results and Discussions}
\label{sec:results}
\input{content/4-results}

\section{Conclusions}
\label{sec:results}
\input{content/5-conclu}

\section{Acknowledgements}
\label{sec:acknow}
\input{content/6-acknow}

\vfill\pagebreak


\bibliographystyle{IEEEbib}
\bibliography{strings,refs}

\end{document}

%% file: content/0-abstract.tex
End-to-end multi-talker speech recognition has garnered great interest as an effective approach to directly transcribe overlapped speech from multiple speakers.
Current methods typically adopt either 1) single-input multiple-output (SIMO) models with a branched encoder, or 2) single-input single-output (SISO) models based on attention-based encoder-decoder architecture with serialized output training (SOT).
In this work, we propose a Cross-Speaker Encoding (CSE) network to address the limitations of SIMO models by aggregating cross-speaker representations.
Furthermore, the CSE model is integrated with SOT to leverage both the advantages of SIMO and SISO while mitigating their drawbacks.
To the best of our knowledge, this work represents an early effort to integrate SIMO and SISO for multi-talker speech recognition.
Experiments on the two-speaker LibrispeechMix dataset show that the CES model reduces word error rate (WER) by 8\% over the SIMO baseline. 
The CSE-SOT model reduces WER by 10\% overall and by 16\% on high-overlap speech compared to the SOT model.
Code is available at \textcolor{Blue}{\href{https://github.com/kjw11/CSEnet-ASR}{https://github.com/kjw11/CSEnet-ASR}}.


%% file: content/1-intro.tex
Automatic speech recognition (ASR) aims to transcribe human speech into text.
Thanks to the rapid progress of deep learning, advanced models like Conformer \cite{gulati2020conformer}, RNN-Transducer (RNN-T) \cite{graves2012sequence, rao2017exploring}, and Attention-based Encoder-Decoder (AED) \cite{chorowski2015attention, chan2016listen} have achieved superior performance in single-speaker ASR tasks.
Meanwhile, multi-talker ASR has emerged as an active research area \cite{li2022recent}, seeking to further empower ASR systems to handle more complex conversational scenarios.
In particular, natural conversational speech often contains one or multiple speakers, with varying degrees of overlap. 
These challenges necessitate dedicated models specifically designed to address them.

There have been diverse approaches proposed for multi-talker ASR.
Conventional cascaded systems utilize a speaker separation model as a front-end, followed by a regular ASR model for recognition \cite{wang2018supervised, yoshioka2018multi, qian2018single}.
More recently, end-to-end models have gained attention due to their promising performance.
End-to-end models can be classified into two types: single input multiple output (SIMO) and single input single output (SISO).
SIMO models employ branch-based architectures that internally separate the mixed speech into isolated branches, followed by shared recognition blocks to transcribe different speakers in parallel \cite{tripathi2020end, chang2020end, lu2021streaming, meng2023sidecar, meng23b_interspeech}. 
To align branches with respective speakers, permutation invariant training (PIT) \cite{yu2017permutation, kolbaek2017multitalker} or heuristic error assignment training (HEAT) \cite{lu2021streaming, tripathi2020end} are applied to calculate the ASR loss.
Compared to cascaded systems, SIMO models combine separation and recognition in a unified structure and without separation loss in training.
In contrast, SISO models serialize transcriptions of different speakers into a single stream, relying on auto-regressive AED frameworks with serialized output training (SOT) \cite{kanda2020serialized, kanda2022streaming, yu2022m2met, kanda2020joint}.
Compared to SIMO models that require pre-defined numbers of speakers and branches, SISO models with SOT leverage an attention-based decoder to transcribe speakers in chronological order, which allows for flexibility in the number of speakers.


Despite recent advances, limitations persist for both SIMO and SISO approaches.
SIMO models propagate speakers' encoding through isolated branches, which could cause repeated or omitted transcriptions \cite{kanda2020serialized, liang2023ba}.
Moreover, incorporating SIMO with other structures like AED and RNN-T introduces complexity due to the multiple output streams \cite{chang2020end}.
Meanwhile, SISO models highly rely on attention mechanisms to disambiguate speakers without explicitly modeling separation.
The Lack of prior information for distinguishing speakers' characteristics could cause performance degradation when facing higher overlapped speech.

In this work, we propose a Cross-Speaker Encoding (CSE) network to address the limitations of SIMO approaches. Further, the CSE is integrated with the SOT strategy to leverage the advantages of SIMO and SISO while mitigating their drawbacks. Specifically, we attribute the drawbacks of SIMO to isolating speakers into separate branches. This “one-time deal” precludes the possibility of different branches conditioning on each other, and impedes potential inter-dependencies between speakers. In our proposed CSE network, a \emph{cross-encoder}, in conjunction with a \emph{joint-HEAT} module, is employed to jointly encode cross-speaker representations. The cross-encoder enables separate branches to condition on each other, while the joint-HEAT simultaneously improves the single-talker performance of the original HEAT and converges the model outputs into a uniform stream. Building on the CSE architecture, we further introduce CSE-SOT as the first attempt to integrate these two methods, reflecting a novel combination of their strengths.
We conducted experiments on simulated conversational speech with varying speaker numbers and overlap degrees.
Results demonstrate that CSE outperforms the branch-based baseline with lower complexity.
Additionally, CSE-SOT significantly surpasses the SOT model while retaining the capability to generalize to more speakers than that in the training data. 

%

\begin{figure*}[htbp]
\begin{center}
\includegraphics[width=1\textwidth]{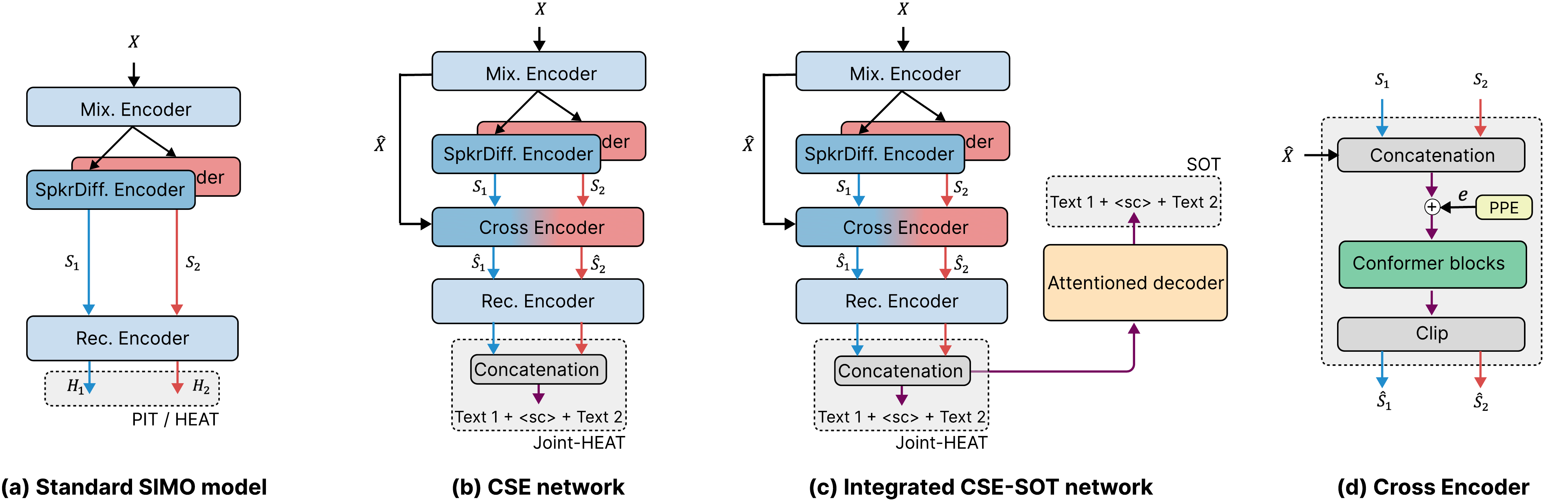}
\end{center}
\vspace{-14pt}
\caption{The architecture of (a) standard branch-based SIMO model, (b) proposed Cross Speaker Encoding (CSE) network, (c) Integrated model with CSE and serialized output training (SOT), and (4) cross-encoder. \textit{SpkrDiff.} refers to speaker differentiator, \textit{Rec.} refers to recognition, and \textit{PPE} refers to partition-wise positional embedding. $\oplus$ stands for concatenation and $\hat{X}$ stands for mixture encoding.}
\vspace{-10pt}
\label{fig:CSE}
\end{figure*}

%% file: content/2-method.tex

In this section, we will first review SIMO models and discuss the limitations, then introduce the proposed Cross-Speaker Encoding (CSE) network.  
For clear demonstration, we focus on the \textit{two-speaker case} when explaining our method.

\vspace{-4pt}
\subsection{Branch-based SIMO model}
\vspace{-2pt}
The branch-based SIMO model provides a unified framework for joint speech separation and recognition.
As shown in Figure \ref{fig:CSE} (a), given a mixture speech feature $X$ and ground truth transcripts $(y^1, y^2)$ for two speakers, the model first encodes $X$ with a mixture encoder.
A separation module with two branches then generates separated representations $S_1$ and $S_2$ for each speaker. 
This separation module can employ either speaker differentiator (SpkrDiff) encoders to encode individual speaker, or masking encoders to mask out unwanted speaker(s).
The separated representations are then fed to a shared recognition encoder, which serves as a standard ASR encoder to predict transcripts separately for each of the two branches.
A key challenge is associating branch output $(H_1, H_2)$ with corresponding target labels $(y^1, y^2)$.
Permutation Invariant Training (PIT) \cite{yu2017permutation} addresses this by permuting all mappings and picking the one with minimum loss to update the model, i.e.,
\begin{equation} \label{eq:3}
\begin{aligned}
\mathcal{L}_{pit}(H_1, H_2, y^1, y^2) &= min(\mathcal{L}_{asr}(H_1, y^1)+\mathcal{L}_{asr}(H_2, y^2), \\
&\mathcal{L}_{asr}(H_1, y^2)+\mathcal{L}_{asr}(H_2, y^1))
\end{aligned}
\end{equation}
where $\mathcal{L}_{asr}$ can be any ASR loss such as connectionist temporal classification (CTC) \cite{graves2006connectionist}. 
The PIT method does not assume any prior knowledge of the mixing conditions, making it applicable to all cases, including fully mixed speech with limited evidence for identifying speakers. 
As a promising alternative to PIT,
Heuristic Error Assignment Training (HEAT) has been explored in \cite{tripathi2020end, lu2021streaming}.
Based on the chronological appearance of speakers, HEAT directly assigns the speaker order to simplify the complexity of PIT during training.
Hence the loss can be calculated by:
\begin{equation} \label{eq:4}
\begin{aligned}
\mathcal{L}_{heat}(H_1, H_2, y^1, y^2) = \mathcal{L}_{asr}(H_1, y^1)+\mathcal{L}_{asr}(H_2, y^2)
\end{aligned}
\end{equation}
For instance, in the two-speaker overlapped speech scene, one branch will be assigned to always transcribe the first-talking speaker, while another branch for the latter speaker.
This strategy has shown to be superior to PIT in streaming multi-talker ASR systems \cite{lu2021streaming}.

\vspace{-4pt}
\subsection{Cross Speaker Encoding network}
\vspace{-2pt}
\label{sec:CSE}
The limitations of SIMO models have been discussed in prior studies \cite{li2022recent, liang2023ba}.
First, the separate encoding branches propagate errors monotonically throughout the model layers.
Hence separation error in early layers could persist to recognition encoders, yielding repeated and omitted transcriptions.
Second, as SIMO models output multiple streams, incorporating SIMO models into frameworks such as AED and  RNN-T incurs extra computational cost \cite{chang2020end}.
We attribute the drawbacks of SIMO to isolating speakers into separate branches. This “one-time deal” precluded the possibility of different branches conditioning on each other, and impeded potential inter-dependencies between speakers.
Additionally, outputting separate branches misaligns with the single-stream manner of common ASR architectures.
Addressing these two points, we proposed a Cross-Speaker Encoding (CSE) model comprising two improvements: \emph{cross-encoder} and \emph{joint-HEAT}.

\noindent \textbf{Cross Encoder}. 
Cross-encoder was proposed to model inter-speaker dependencies, illustrated in Figure \ref{fig:CSE} (b) and (d).
It comprises four steps: 1) concatenating the outputs of separated branches $(S_1, S_2)$ and mixture encoding $\hat{X}$ into a joint encoding $[\hat{X}, S_1, S_2]$.
2) Adding this joint encoding with a learnable partition-wise positional embedding $e$, where frames belonging to the same partition share the same positional embedding.
3) Feeding the derived representations into Conformer blocks \cite{gulati2020conformer},  where the self-attention layer provides a global view, allowing the branches to attend to each other mutually.
This enables omission errors could be compensated from mixture encoding, and repetitions could be censored based on other branches.
Finally, the joint encoding was clipped as $\hat{S_1}$ and $\hat{S_2}$, i.e., encoded versions of $S_1$ and $S_2$ with additional context, allowing the shared recognition encoders to generate respective transcriptions. 
Note that, one can directly feed concatenated encoding into recognition encoders without clipping.
However, our preliminary experiments show this cannot bring additional performance gain while introducing considerable computational cost due to the quadratic complexity of self-attention.

\noindent \textbf{joint-HEAT}.
We introduce joint-HEAT as a straightforward solution to unify separate output streams.
First, we concatenated outputs of different branches, then adopted HEAT loss to disambiguate the labels based on the speaking order, as used in \cite{lu2021streaming}.
Specifically, during training we concatenation text labels of multi-talker speech as $[y^1,$$<$sc$>$$,y^2]$, where $y_1$ associates with the first-talking speaker and $y_2$ associated with the second one.
The $<$sc$>$ token indicates speaker-changes boundaries between texts from separate speakers.
Joint-HEAT can also improve single-talker performance compared to the original HEAT.
Our preliminary studies show that in single-speaker speech, models trained with HEAT produce omission errors at the end of sentences.
This may be because the separation module attempts to only model the "first speaking speaker", hence omitting part of the tokens.
Concatenating the outputs makes the predictions consider both branches, empirically alleviating this problem.

\vspace{-4pt}
\subsection{Integrated CSE-SOT model}
\vspace{-2pt}
As joint-HEAT unified separate output streams, we explored a hybrid SIMO-SISO system by employing an attentioned decoder with a CSE encoder and used SOT to guide the decoder training.
The integrated CSE-SOT model complements the weaknesses of each method when used alone.
First, the attention decoder can better handle speech context and temporal dependencies compared to sole SIMO models.
Then SIMO models explicitly model speaker separation, facilitates speaker disambiguation for the SISO decoder.

\vspace{-3pt}

%% file: content/3-settings.tex
\vspace{-4pt}
\subsection{Dataset}
\vspace{-2pt}
We use \textit{LibriSpeechMix} (LSM) \cite{kanda2020serialized} as a benchmark dataset in our experiments.
This dataset is simulated from 960-hour \textit{LibriSpeech} (LS) \cite{panayotov2015librispeech} corpus with 2-speaker (LSM-2mix) and 3-speak conditions (LSM-3mix).
Since LibriSpeechMix only provides standard development and test sets, we simulated a 2-speaker training set following the same protocol as in \cite{kanda2020serialized}.
Specifically, we randomly sample two utterances from LS training set with random delay offset and speed perturbation between $\{0.9,1.0,1.1\}$ times.
We combine the original LS training set and this simulated data, then subset 400k utterances (1.7k hours) for efficient training.

For a realistic evaluation, we expect a multi-talker ASR system to be able to handle both single-talker and multi-talker speech. 
Therefore, we utilized both the LS and LSM datasets to examine all models.
Furthermore, given the diverse overlap conditions within the test set, we partitioned the LSM test set into three subsets, denoting them as \textit{low-overlap}, \textit{median-overlap}, and \textit{high-overlap} scenarios respectively. 
The corresponding overlap ratios are bounded by (0, 0.2], (0.2, 0.5], and (0.5, 1.0]. 
The overlap ratio here is defined as the number of overlapped frames divided by the total number of frames.

\vspace{-4pt}
\subsection{Model settings}
\vspace{-2pt}
We implement models based on the ASR Conformer encoder using ESPnet2 \cite{watanabe2018espnet} toolkit.
We used 80-dimensional Mel-filter bank as the input feature with the speed perturbation described above.
For the SIMO model, apart from the original convolutional subsampling layer, we used the same CNN layer as the Mix encoder, and 4 Conformer blocks as SpkrDiff encoder for each branch, finally 8 Conformer blocks as shared recognition encoder.
Therefore there are 16 conformer blocks, and each Conformer block has a 4-head self-attention with 256 hidden units and two 1024-dimensional feed-forward layers (macaron style).
On top of the SIMO model, the CSE model used the same Mix and SpkrDiff encoder, but 2 Conformer blocks as cross-encoder and 6 Conformer blocks as recognition encoder.
Therefore SIMO CSE models have the same number of parameters of 33.20M.
As for the SOT baseline, a conformer encoder with 16 blocks was adopted followed by an 8-block transformer decoder.
Each transformer block comprises a self-attention layer with 4 attention heads and 256 hidden units, but a 2048-dimensional feed-forward layer.
The CSE-SOT model used the same encoder as the CSE model and the same decoder as the SOT model.
As a result, both CSE-SOT and SOT models have 45.24M parameters.

\vspace{-4pt}
\subsection{Training settings \& Metrics}
\vspace{-2pt}
For all of our experiments, we trained the models with 35 epochs and fused the best 10 epochs on the dev set as final models.
We use Adam optimizer with a learning rate of $5e-4$ and warm-up steps of $25,000$.
For SOT and CSE-SOT models, we use joint CTC/attention \cite{watanabe2017hybrid} as the training objective, with a CTC weight of $0.3$.
For single-speaker cases, we evaluated using the standard word error rate (WER). For multi-talker samples, we applied the permutation-invariant WER as in \cite{kanda2020serialized}, which chooses the speaker order with the lowest WER for scoring. 
However, we observed that the overall WER was dominated by samples with only mild overlaps (i.e., over 40\% samples have overlap ratios $\leq$ 0.2). Therefore, we further employed an overlap-aware WER (OA-WER) that averages WERs across different overlap ratios, in order to assess the models' ability to handle different degrees of overlap.

\vspace{-3pt}

\input{tables/t1}

%% file: tables/t1.tex
\begin{table*}[h]
\centering

\caption{
Performance comparison (WER\%) of different systems on Librispeech and LibrispeechMix-2mix evaluation set. The Test (Conditional) set involves three subsets with different overlap ratios. 
For simplification, we denote these three subsets as low, median, and high-overlap in the text. OA-WER refers to the averaged results across these three subsets.
}
\vspace{3pt}
\label{tab:all-wer}
\scalebox{0.9}{
\begin{tabular}{l|l|cc|cc|cccc}
\bottomrule

\multirow{3}{*}{\textbf{ID}} & \multirow{3}{*}{\textbf{System}}  &\multicolumn{2}{c|}{\textbf{Librispeech}} &\multicolumn{6}{c}{\textbf{LibrispeechMix-2mix}} \\
 \cline{3-4} \cline{5-10} 
& & \multirow{2}{*}{\textbf{dev-clean}} & \multirow{2}{*}{\textbf{test-clean}} & \multirow{2}{*}{\textbf{Dev}} & \multirow{2}{*}{\shortstack{\textbf{Test}\\(Overall)}}& \multicolumn{4}{c}{\textbf{Test} (Conditional)} \\
\cline{7-10}
& & & & & & (0, 0.2] & (0.2, 0.5] & (0.5, 1.0] & OA-WER \\
\noalign{\hrule height 0.9pt}
A1 & SIMO w/ PIT & \textbf{6.9} & \textbf{6.8} & 12.3 & 11.6 & 8.9 & 12.3 & 17.4 & 12.8\\
A2 & SIMO w/ HEAT & 9.7 & 9.9 & \textbf{12.0} & \textbf{11.1} & \textbf{8.3} & \textbf{11.7} & \textbf{16.5} & \textbf{12.2}\\
\hline
B1 & SIMO w/ Joint-HEAT & 7.7 & 7.1 & 12.0 & 11.2 & 8.8 & 11.7 & 16.8& 12.3\\
B2 & CSE & \textbf{6.9} & \textbf{6.8} & \textbf{11.8} & \textbf{10.7} & \textbf{8.2} & \textbf{11.3} & \textbf{16.1} & \textbf{11.9}\\
\hline
C1 & \quad - PPE & \textbf{6.9} & 6.8 & 11.9 & 11.0 & 8.4 & 11.7 & 16.8 & 12.3\\
C2 & \quad - mix. encoding & \textbf{6.9} & \textbf{6.7} & 11.9 & 10.9 & 8.3 & 11.7 & 16.4& 12.1 \\
\hline
D1 & SOT & \textbf{4.2} & \textbf{5.4} & 9.5 & 9.4 & 7.3 & 9.9 & 13.9 & 10.3\\
D2 & CSE-SOT & 4.5 & \textbf{5.5} & \textbf{8.1} & \textbf{8.4} & \textbf{7.2} & \textbf{8.3} & \textbf{12.0} & \textbf{9.2}\\

\toprule
\end{tabular}
}
\vspace{-0.6cm}
\end{table*}

%% file: content/4-results.tex
\vspace{-4pt}
\subsection{Results of CSE}
\noindent \textbf{PIT \textit{vs}. HEAT}.
First, we compared the baseline SIMO approach, as depicted in Figure \ref{fig:CSE} (a), with either PIT or HEAT loss.
Shown in Table \ref{tab:all-wer}, the HEAT model (system A2) is significantly worse than the PIT model (system A1) in single-talker cases.
As discussed in Section \ref{sec:CSE}, our experiments observed that the HEAT model frequently produces omission errors at the end of sentences.
This may be because, in our setting, HEAT models only use one branch output to transcribe single-talker speech, without any constraint on another branch to suppress potential token leakage.
In contrast, on LibrispeechMix-2mix (LSM-2mix) multi-talker set, the HEAT model shows superior accuracy than PIT, this aligns with the observations in \cite{lu2021streaming}.
This improvement is consistent across different overlap ratios, especially for high-overlap samples ($16.5$ \textit{vs.}$17.4$).
We argue that expressly assigning one SpkrDiff encoder to capture one exact speaker (e.g., the first-talking one) can well guide this encoder to learn specific patterns.

\noindent \textbf{Proposed methods}.
In system B1 of Table \ref{tab:all-wer}, we first replace HEAT with our proposed joint-HEAT.
We can see the performance on single-talker utterances was boosted by a large margin as expected, while the performance on multi-talker speech is equivalent ($12.2$ \textit{vs.}$12.3$ on OA-WER).
Enhanced from system B1, system B2 demonstrates the effectiveness of the suggested CSE model, which is further equipped with a cross-encoder.
For single-talker cases, the CSE model further improves the performance and fills the gaps in comparison to the PIT model.
For multi-talker cases, the introduction of a cross-encoder leads to additional improvement, possibly due to the benefit of information sharing between the two branches.
We will discuss this hypothesis in the visualization section below.

\noindent \textbf{Ablation study on Cross-Encoder}.
To validate the effectiveness of the two components in cross-encoder, we conducted ablation studies by removing internal components. First, system C1 removes partition-wise positional embedding (PPE). 
Without PPE, the cross-encoder won't be explicitly instructed on which frames belong to which partition, hence solely relying on distance information from relative positional encoding.
This leads to considerable performance decline, especially for high-overlap speech ($16.8$ \textit{vs.}$16.1$).
System C2 removed mixture encoding from joint encoding (i.e., only concatenate representations of two branches). 
In this case, the information omitted by both branches may not be recovered.
This could explain the consistent performance degeneration on multi-talker speech.

\vspace{-4pt}
\subsection{Results of CSE-SOT}
\noindent \textbf{CSE-SOT vs. SOT}.
We compared the baseline SOT model with the proposed CSE-SOT model shown as system D1 and D2 in Table \ref{tab:all-wer}.
It is not surprising that CSE-SOT did not show improvement on single-talker cases, since SOT inherits a standard AED architecture designed for single-talker ASR.
For multi-talker cases, CES-SOT shows equivalent performance on low-overlap speech ($7.2$ \textit{vs.}$7.3$), while showing remarkable improvement on both median and high-overlap scenarios.
In particular, compared to the SOT baseline, CSE-SOT model attains $16.2\%$ of relative improvement ($8.3$ \textit{vs.}$9.9$) on median-overlap speech, and $13.7\%$ of relative improvement ($12.0$ \textit{vs.}$13.9$) on high-overlap speech.
These findings indicate that SOT, relying solely on cross-attention, has limitations in transcribing overlapping speech beyond minor conditions.
In contrast, the proposed CSE-SOT, serving as a hybrid SIMO-SISO framework, explicitly models separation with a SIMO structure and offers a straightforward solution to mitigate the above drawback.

\input{tables/t2}

\noindent \textbf{Generalize to more speakers}.
One benefit of the SOT model is its ability to generalize to more speakers beyond that in training data.
We also evaluated the CSE-SOT model under this condition.
As shown in Table \ref{tab:3spkr}, it demonstrated that the CSE-SOT model still retains the same ability even though there are only two branches in the CSE module.
Note that the results of CSE-SOT are slightly worse than the SOT baseline.
A possible reason is one of the branches encoded two speakers simultaneously, whereas this may trouble the decoder to further distinguish these two.
More investigation will be conducted in our future work for more insight into this phenomenon.

\vspace{-4pt}
\subsection{Visualization}
To better understand the CSE model, we investigated the self-attention layer in the Conformer block of the cross-encoder.
Figure \ref{fig:att1} illustrates attention matrices of 4 attention heads from the last Conformer block, where each row represents a weight vector that indicates how outputs attend.
The visualizations show that different attention heads have distinct roles. A certain head attends to frame-level cues (diagonal matrices in head (a)) while another focuses on partition-level details (dense matrices in head (b)). Heads (c) and (d) exhibit the combination of both patterns. Interestingly, head (a) shows S1 and S2 mutually attending -- the output of $\hat{S}_2$ mainly focused on $S_1$, while $\hat{S}_1$ focused on $\hat{S}_2$ -- implying two branches may be swapped, \emph{this serves as a similar function of PIT but without extra training complexity}.


%% file: tables/t2.tex
\vspace{-10pt}
\begin{table}[htbp]
\centering
\caption{Performance comparison (WER\%) between the SOT baseline and CSE-SOT model on the LibrispeechMix-3mix evaluation sets. Both models are train on single-talker plus 2-talker data, and evaluated on 3-talker data.}
\vspace{3pt}
\label{tab:3spkr}
\scalebox{0.9}{
\setlength{\tabcolsep}{4pt}
\begin{tabular}{l|cc|cccc}
\bottomrule

\multirow{2}{*}{\textbf{System}} & \multirow{2}{*}{\textbf{Dev}} & \multirow{2}{*}{\shortstack{\textbf{Test} \\ (Overall)}}& \multicolumn{4}{c}{\textbf{Test} (Conditional)} \\
\cline{4-7}
 &  &  & (0,20] & (20,50] & (50,100] & OA-WER \\
\noalign{\hrule height 0.9pt}
SOT & \textbf{24.2} & \textbf{24.3}        &  \textbf{18.1}           & \textbf{24.0}             &   \textbf{31.0}             & \textbf{24.3} \\
\hline
 CSE-SOT & \textbf{24.2} & 24.5 & \textbf{18.1} & \textbf{24.1} & 31.8 & 24.7 \\

\toprule
\end{tabular}
}
\end{table}
\vspace{-5pt}

%% file: content/5-conclu.tex
\begin{figure}[tbp]
\begin{center}
\includegraphics[width=.9\linewidth,scale=1.00]{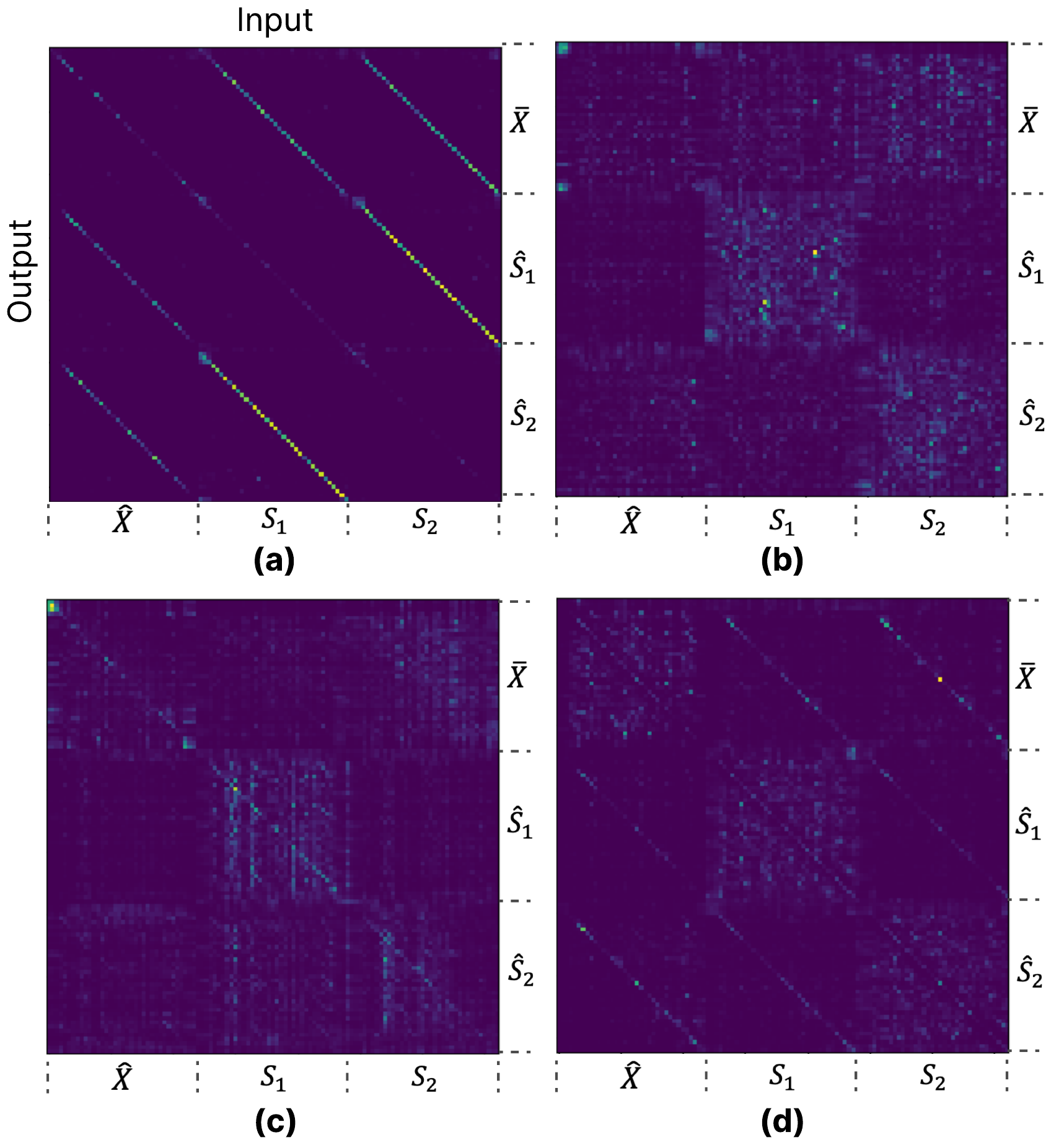}
\end{center}
\vspace{-0.7cm}
\caption{
Attention matrices of the last conformer block in the cross-encoder. (a)-(d) refers to 4 attention heads.
$\hat{X}$, $S_1$ and $S_2$ denote the input partitions. $\bar{X}$, $\hat{S}_1$ and $\hat{S}_2$ denote the corresponding output partitions.
}
\vspace{-0.6cm}
\label{fig:att1}
\end{figure}

\vspace{-2pt}
In this work, we discussed the limitations of commonly used branch-based SIMO models for the multi-talker ASR task, then proposed a cross-speaker encoding (CSE) network consisting of a cross-encoder and joint-HEAT module.
Experiments validated that the cross-encoder allows separate branches to condition on each other, while the joint-HEAT simultaneously enhances the single-talker performance of the original HEAT and converges the model outputs into a uniform stream.
Further, the CSE is integrated with the SOT strategy to leverage both the advantages of SIMO and SISO while mitigating their drawbacks.
Compared to the SOT baseline, the integrated CSE-SOT model reduces WER by 10\% overall and by 16\% on high-overlap speech, demonstrating promising potential for further investigation.
\vspace{-4pt}

%% file: content/6-acknow.tex
\vspace{-4pt}
This work is supported by the HKSARG Research Grants Council’s Theme-based Research Grant Scheme (Project No. T45- 407/19N) and the CUHK Stanley Ho Big Data Decision Research Centre.